\documentclass[aps,prl,preprint,showpacs,superscriptaddress,amsmath,amssym,nofootinbib]{revtex4}
\usepackage{graphicx}
\usepackage{dcolumn}
\usepackage{bm}
\usepackage{color}
\usepackage{xspace}
\usepackage{comment}
\usepackage{dcolumn}
\usepackage{bm}
\usepackage{longtable}
\usepackage{xspace}
\begin{document}

\title{Enhancement of the Triple Alpha Rate in a Hot
 Dense Medium}
\author{Mary Beard\footnote{Deceased}}
\affiliation{Department of Physics, University of Notre Dame, Notre  Dame, Indiana 46556, USA\\
Joint Institute for Nuclear Astrophysics}
\author{Sam M. Austin\footnote{austin@nscl.msu.edu}}
\author{Richard Cyburt\footnote{Presently at Concord University, Athens, WV; rcyburt@concord.edu}}
\affiliation{National Superconducting Cyclotron Laboratory\\
Joint Institute for Nuclear Astrophysics\\
Michigan State University, 640 South Shaw Lane, East Lansing, MI 48824-1321, U.S.A.}

\begin{abstract}

In a sufficiently hot and dense astrophysical environment the rate of the triple-alpha ($3\alpha$) reaction can  increase greatly over the value appropriate for helium burning stars owing to hadronically induced de-excitation of the Hoyle state. In this paper we use a statistical model to evaluate the enhancement  as a function of temperature and density. For  a density of 10$^6$ gm cm$^{-3}$ enhancements can exceed a factor of one-hundred. In high temperature/density situations, the enhanced $3\alpha$~rate is a better   estimate of this rate and should be used in these circumstances.   We then examine the effect of these enhancements on production of $^{12}$C in the neutrino wind following a supernova explosion and in an x-ray burster.

\end{abstract}
\date{\today}

\pacs{26.20Fj;26.30.-k;25.40.-h}

\maketitle

\begin{figure}
\includegraphics[width=4cm]{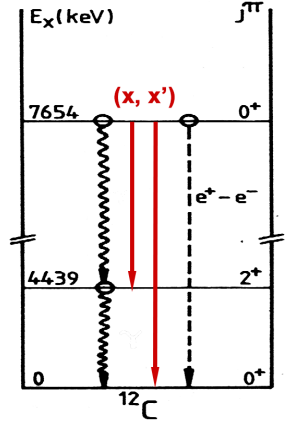}
\caption{(Color on line). The radiative width for typical helium burning stars results from the gamma ray cascade from the Hoyle state through the the 4.44 MeV $2^+$ state and the (much weaker) pair decay of the Hoyle state.  The additional contributions we estimate, shown as solid downward arrows, are mediated by inelastic scattering of protons, neutrons and alpha particles leading from  the Hoyle state to the $2^+$ state and to the ground state.\label{fig:3alpha}}
\end{figure}

 The triple alpha  ($3\alpha$) process that converts $^4$He into $^{12}$C is one of the fundamental reactions in astrophysics; its rate influences the stellar production of many elements \cite{her06,tur07,wes13}. For the stellar conditions typically encountered in helium burning stars, the $3\alpha$ rate  is  proportional to  the radiative width of the 7.65 MeV $0^+$ state (the Hoyle state) in $^{12}$C  and is known to within about 10\%. This will be true whenever the triple-alpha reaction proceeds through resonant processes.  Recent theoretical calculations, \cite{ngu13,sun16} have  shown that this is the case for $T>10^8 K$.  At large values of the temperature ($T$) and  density ($\rho$), however, the width of the Hoyle State is increased by particle induced de-excitation leading either to the ground state or to the first excited $2^+$ state of $^{12}$C at 4.44 MeV as shown in  Fig.~\ref{fig:3alpha}. This increases (enhances) the rate of the $3\alpha$~process. A principal motivation of this paper is to determine these enhancements using presently available techniques and investigate whether they might be large enough to influence other astrophysical phenomena that occur at high $T$ and $\rho$.  For example, might the enhanced rates produce sufficient seeds in the  neutrino driven wind of a  core-collapse supernovae to make a successful r process less likely.

These enhancement processes have been studied theoretically in the past. Shaw and Clayton \cite{sha67} examined electromagnetic effects: Coulomb excitation by alpha particles and electron induced processes. The effects were found to be much smaller than those for the nuclear reaction induced effects considered here, and unimportant for densities less than $10^9$ g cm$^{-3}$. They will not be discussed further in this paper. Truran and Kozlovsky \cite{tru69} considered nuclear induced processes but before there were reliable measurements or estimates of the relevant cross sections.

Following these theoretical estimates, experimental studies of inelastic proton \cite{dav71} and alpha scattering \cite{mor70} from the ground state of $^{12}$C to the Hoyle state were carried out. The corresponding enhancements were calculated from the inverse rates that correspond to these cross sections. These preliminary studies indicated that the enhancements could be significant at the temperatures and densities encountered in supernovae. There were, however,  no reliable estimates of cross sections for neutron inelastic scattering that because of the absence of Coulomb effects would be expected to dominate the enhancements. Nor could experiments yield estimates of inverse rates for processes that lead from the Hoyle state to the first excited $2^+$ state in $^{12}$C.  Such cross sections are not experimentally measurable and must be obtained from theoretical estimates.

In this paper we attempt to deal with these deficiencies. Although any particle present can cause an enhancement,  the particle densities and temperatures required for significant enhancements are large, so that in practice it is necessary to consider only the effects of neutrons, protons, and alpha particles.

We provide the relevant reaction rate background; describe the experimental inelastic cross sections  and the theoretical calculations used to generate cross sections not available from experiment; and present the enhancements obtained as a function of temperature at a density of $10^6$ g cm$^{-3}$. Since the enhancements are directly proportional to the density, these results are sufficient for  applications to astrophysical phenomena. Finally we present preliminary estimates of the effects of the enhancements  in astrophysical applications and discuss the experimental and theoretical advances that could improve the accuracy of these initial results.

The procedures used follow those described in Davids and Bonner \cite{dav71}. For the two body reactions induced by neutrons, the reaction rate of $^{12}$C with number density $N_{^{12}C}$, and neutrons with number density $N_n$, is given by
\begin{equation}
r=N_nN_{^{12}C}<\sigma v>   cm^{-3} sec^{-1}
\end{equation}
where $\sigma$ is the total reaction cross section for exciting the Hoyle state from the initial state (g.s. or 4.44 MeV $2^+$ state), $v$ is the relative velocity, and the average is over the Maxwellian velocity distribution of the two species. For inelastic neutron scattering on $^{12}$C
\begin{equation}
<\sigma v>_{nn\prime}=\left(\frac{8}{\pi\mu}\right)^{1/2}\left(\frac{1}{kT}\right)^{-3/2}\int^{\infty}_0 E^\prime\sigma_{n,n'}(E^\prime)exp(-E^\prime/kT)dE^\prime
\end{equation}
For our purposes we need the inverse of the reaction exciting the Hoyle state, namely
\begin{equation}
<\sigma v>_{n^\prime n}=\left(\frac{2I+1}{2I^\prime+1}\right)exp(-Q/kT)<\sigma v>_{nn^\prime}
\end{equation}

Here $I$ and $I^\prime$ are the spins of the initial and final states for forward excitation of $^{12}$C: $0^+$ and  $0^+$ for the ground state to Hoyle state transition, and $2^+$ and $0^+$ for transitions from the 4.44 MeV $2^+$ state to the Hoyle state, resp. $Q=$ ~-7.654 (-3.215) for
excitation of the Hoyle state, from the g.s.($2^+$ state). The life time for inelastic neutron de-excitation is
\begin{equation}
\tau_{n^\prime n}(^{12}C^{Hoyle})=(N_n<\sigma v>_{n^\prime n})^{-1}  sec.
\end{equation}
We define  $R$ as the ratio of the radiative lifetime to the particle-induced lifetime.
\begin{equation}
R=\tau_\gamma/\tau_{n^\prime n}=\tau_\gamma N_n<\sigma v>_{n^\prime n}.
\end{equation}
Inserting the experimental value $\tau_\gamma=1.710\times10^{-13}$sec \cite{fre14,aus12}, the values of the relevant constants, and  expressing the energy as kinetic energy above threshold, one obtains
\begin{equation}
R=k_n\rho_n T_9^{-1.5}
 C_{spin} \int^{\infty}_0\sigma_{nn\prime}(E)(E-Q)exp(-11.605E/T_9)dE
\end{equation}
where $E$ is the c.m. energy above threshold, $\rho_n$ is the neutron density in g cm$^{-3}$, $T_9=T/10^9$,
and $\sigma_{nn\prime}(E)$ is the cross section in mb. For transitions to the Hoyle state from the ground state (4.44 MeV  2$^+$ state),  C$_{spin}$ = 1(5).

In all these equations, for proton inelastic scattering substitute $p$ for $n$ and $p^\prime$ for $n^\prime$; for alpha particle inelastic scattering substitute $\alpha$ for $n$ and $\alpha^\prime$ for $n^\prime$. The multiplying constants are: k$_n=6.557 {\times 10^{-6}}$; k$_p= 6.565 {\times 10^{-6}}$; k$_\alpha=9.200 {\times 10^{-7}}$.

Experimental values are available in the literature for a few of the inelastic cross sections, but they are sparse and often do not extend low enough in energy toward the reaction thresholds. Most important, for the neutron induced reactions expected to dominate at relatively low temperatures there are no results in the relevant energy range of up to a few MeV above threshold. Nor are there estimates, for any projectile, of the cross sections from the 4.44 MeV state to the Hoyle State.

For the important energies near threshold, compound nuclear processes are expected to dominate and one might first consider employing an R-matrix description. Unfortunately, a large number of levels, some narrow and some broad, influence these cross sections \cite{ajz91,hal14} so that any analysis will be complex, and will, at the moment, lack the necessary experimental information. The best of the available analyzes \cite{hal14}, presently does not extend to the compound nucleus energies we require, partially because data is insufficient or contradictory. It appears that significant improvements in the R-matrix approach will take significant effort and time \cite{deb17}.  Many of these comments apply to other possible approaches.

The best available option is  to follow the standard approach in astrophysics (see, for example, a description of the JINA REACLIB database \cite{cyb10}), of obtaining unmeasured compound nuclear cross sections from the statistical Hauser-Feshbach (HF)  model~\cite{wol51,hau52}. For this purpose we use the reaction code TALYS (version 1.8)~\cite{tal07}, a widely accepted modern implementation of the HF model.  The underlying principle of this  statistical model picture is that the interaction of a target and projectile result in the formation of a compound nucleus at a sufficiently high excitation energy that individual nuclear levels can be treated in an average manner and that the system is fully equilibrated before it decays into the final reaction channels. The probabilities for the creation and decay of the compound nucleus are expressed in terms of the transmission functions for its formation and break up.  For particle channels, the transmission functions are obtained from optical model calculations. Aside from the transmission functions, one requires level densities, width fluctuation corrections and other descriptive details of the target.

For this light system one cannot expect a priori that the basic assumptions outlined above are  well fulfilled; we use the HF approach because there are no realistic alternatives. A related uncertainty lies in the choice of a particular optical potential for the HF calculations. We used the default models of TALYS (version 1.8) \cite{tal07}: for protons and neutrons, a global and local potential based on the Koning and Delaroche model~\cite{kon03}; and for alpha particles, the potentials of Avrigeanu, {\it et al.}~\cite{avr14}.

\begin{figure}
\includegraphics[width=8.4cm]{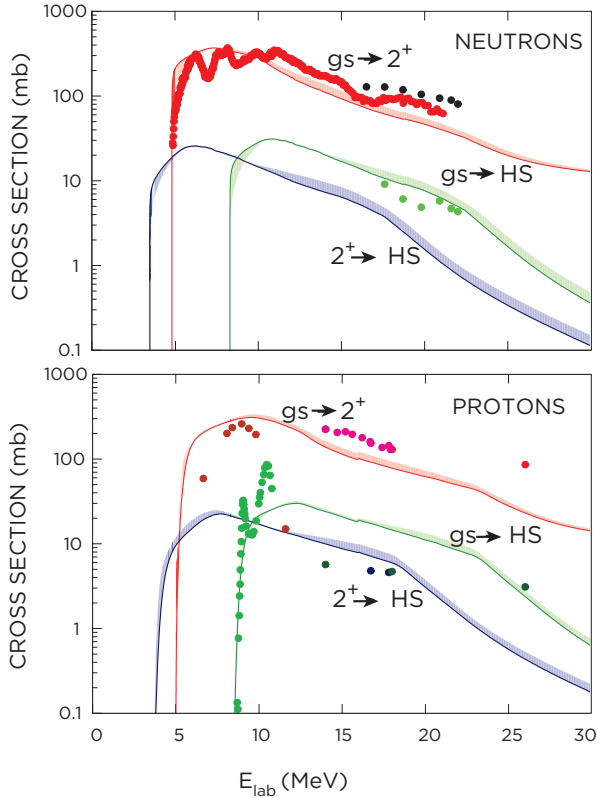}
 \caption{(Color on line) Inelastic scattering cross sections for neutrons (top panel) and protons (bottom panel). Results for the default OMPs described in the text are shown as solid lines, and the maximum and minimum cross sections for other potentials noted in the text are included within the shaded areas. For each projectile results are shown for scattering leading from: the ground state to the 4.44 MeV $2^+$ state, the ground state to the 7.65 MeV Hoyle State, and the 4.44 MeV $2^+$ state to the Hoyle state. Available experimental results are shown as discrete points, for neutrons from \cite{ols89,rog75}, for protons from \cite{dav71,pee57,bar66,har99}. \label{fig:crosssections}}
\end{figure}

To evaluate the resulting uncertainties in a conservative manner,  we calculate all cross sections in a systematic fashion, using the TALYS default parameter values, compare the results to the  available data and, thereby, assess the reliability of the model. Cross sections were also calculated for n, p and $\alpha$ inelastic reactions using different optical model parameters (three for protons and neutrons and nine for alpha particles, as cited in \cite{tal07}) The default results and the spread of results for the other  models  are shown in Fig.~\ref{fig:crosssections}. Although they are not directly relevant to our enhancement calculations, we also show the cross sections for the transition from the ground state to the $2^+$ state at 4.44 MeV since more experimental data are available for this strong transition.

Cross sections calculated with the various OMPs differ by less than 30\% up to approximately 20~MeV; cross sections within 2~MeV of threshold generally dominate the enhancements.  The calculations generally lie within about a factor of two to three of the sparse experiential data, sometimes higher and sometimes lower; the energy dependence of the cross sections is generally reproduced.  The single exception is for the resonance in proton scattering to the Hoyle state. This overall level of agreement is similar to that obtained for heavier nuclei.

The energy dependencies of the cross sections for neutrons have  well known behaviors that differ from those of the charged particles because of the absence of a coulomb barrier. The forward, endothermic, cross sections shown here vary approximately as $E^{\prime1/2}$.  The inverse, exothermic, cross sections exhibit the well known $1/E^{\prime1/2}$ or $1/v$ behavior. As we shall see, this can lead, for neutrons, to large $3\alpha$~enhancements even at relatively low  temperatures.

The enhancements were calculated for the inverse of each of the contributing transition rates:  $gs\rightarrow7.65$ (Hoyle), and $4.44\rightarrow7.65$ state for incident neutrons, protons and alpha particles, using equation (6). For the case of the proton inelastic scattering to the Hoyle state we  used the experimental data up to 2.30 MeV; at lower energies the cross section has strong resonances that are not reproduced by the TALYS calculations. Otherwise the default TALYS cross sections were used. The cross sections were (accurately) fitted with cubic splines and the integrals performed with the MathCad routine over the energy range from near threshold to $10$ MeV above threshold.

\begin{figure}
\includegraphics[width=8.4cm]{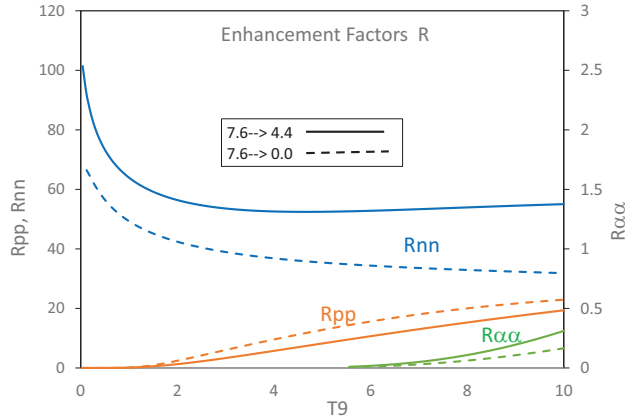}
\caption{(Color on line) Ratios of the rate induced by the indicated transitions to the measured (gamma + pair decay) rate.  The ratios were calculated for a particle density of $10^6$ gm cm$^{-3}$. The alpha ratios are plotted on the expanded scale on the right hand ordinate.
\label{fig:ratios} }
\end{figure}

Most contributions to the enhancements are for energies less than 2 MeV above threshold; except for the small alpha enhancements, all have  converged to within 3\% by 5 MeV, even in the most demanding case: $T_9=10$. Thus, factor of two or three cross section uncertainties at higher energies shown in Fig.~\ref{fig:crosssections}  have small effects on the ratios. The calculated ratios are shown in Fig.~\ref{fig:ratios}.

As expected, the neutron induced enhancements are largest, proton induced enhancements are smaller, and the alpha induced enhancements smaller still. Except for the proton induced transitions, where the larger ground state enhancement owes to the large resonant cross sections at low energies, the enhancements from from 4.44 MeV $2^+$ state to the Hoyle state are larger because of the influence of a spin factor of five.

For applications, it is the sums of the  two cross sections of each projectile that are relevant. We see from  Fig.~\ref{fig:ratios} that these enhancements can be large for sufficiently large  $T_9$ and $\rho$. For neutrons only, enhancements are larger for small $T_9$.  Based on the cross section uncertainties, it seems a fair summary to conclude that the enhancements are known to within about a factor of 2 to 3. Thus for example, the enhancement factor for neutrons at $T_9=1.0$, is 115, the sum of the two values for neutrons shown in Fig.~\ref{fig:ratios}. For a factor of three uncertainty, it would lie between 38 and 345.  Such large enhancements should be taken into account in calculations at high $T$, $\rho$.

\begin{figure}
\includegraphics[width=8.4cm]{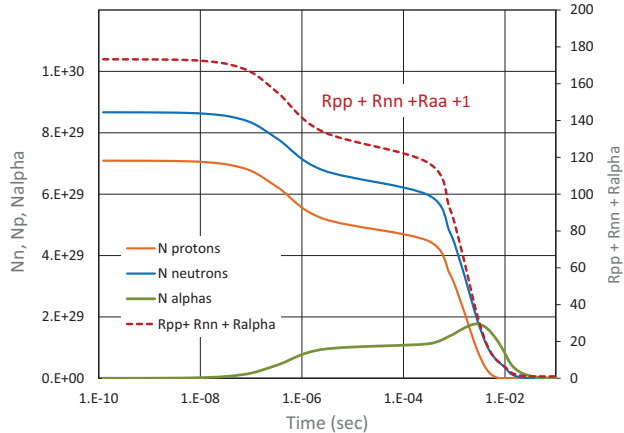}
\caption{(Color on line) On the left hand ordinate are shown, as a function of time, the neutron, proton and alpha particle densities calculated following the adiabatic model. Initially the   temperature was set at $T_9=9$ to insure nuclear statistical equilibrium, the entropy at $S=90k_B$, $Y_e=0.45$, and velocity=7500 km sec$^{-1}$. The calculated overall enhancement, owing to induced de-excitation of the Hoyle state by protons, neutrons and alpha particles is shown on the right hand ordinate.
\label{fig:adiabatic} }
\end{figure}

To investigate the magnitude of these effects in an astrophysical scenario, we  calculated the enhancements for an adiabatic model  \cite{fre99} as implemented by Schatz,{\it et al.} \cite{sch12}.  In this model the initial protons and neutrons are in nuclear statistical equilibrium and are later incorporated into  alpha particles, then into $^{12}$C~and eventually into heavier seeds. The results in Fig.~\ref{fig:adiabatic} show that these enhancements are large.

It has usually been found that in this model the $\alpha\alpha n$~process leading to $^9$Be dominates the flow into $^{12}$C,~but if the $3\alpha$~process is sufficiently enhanced and competes strongly, the overall flow into heavy seeds may increase, leading to a larger number of seeds, a smaller neutron to seed ratio, and a less robust r-process. We have made a preliminary estimate based on the above adiabatic model as summarized in Fig.~\ref{fig:adiabatic}, and find that the enhanced $3\alpha$~rate dominates the production of $^{12}$C, presumably leading to a larger seed abundance. This would remain true if the enhanced rates are a factor of three smaller than calculated here. On the other hand, if there were a strong resonance at low neutron energies, as there is for the proton channel, the enhancements might be still  larger. We intend to investigate these possibilities systematically in future work using more realistic models \cite{jin17}.

Another site where significant enhancements of the $3\alpha$ rate might be important is accreting neutron stars and the resulting x-ray bursts. An increased formation of $^{12}$C  could increase energy generation during the giant outbursts seen in some x-ray bursters. However, the densities and temperatures seen in two possible models of the process \cite{sch17} indicate that the enhancement would reach a maximum of 30\% during the onset of the burst.

One might ask whether the enhancement processes considered here could affect other reaction rates.  Indeed, any process involving gamma decay of a state with a larger particle decay width will have a rate proportional to the radiative width and be susceptible to enhancement. But since  radiative widths are usually large (compared to that of the Hoyle state) enhancements are less likely to be important. That is the case for the $\alpha\alpha n$ process discussed above.

 The situation for the present then appears to be that, in situations where the densities and temperature are large, a reasonable estimate of the triple alpha reaction rate is given by the enhancement factor of Fig.~\ref{fig:ratios}. Uncertainties are probably a factor of three, and the enhancements could presumably be larger or smaller.  If such enhancements cause a significant change in calculated astrophysical phenomena, a significant experimental and  theoretical  effort would then be warranted to better constrain the enhancements.

The neutron cross section from the ground state to the Hoyle state can, in principle, be measured, but the cross sections are relatively small and the measurements will be difficult. If successful, in the context of the present calculations the major uncertainty would then arise from the uncertainties in the input cross sections for the transitions from the $2^+$ state to the Hoyle state where there is no constraint from experiment. Statistical approaches such as those implemented in the TALYS code are typically assumed to be uncertain by factors of two to three, which is consistent with the differences from experiment observed in Fig.~\ref{fig:crosssections}.

A detailed R-Matrix approach, coupled with better understanding of the relevant level structure of the mass-13 nuclei, could probably yield better results, and is highly desirable. But such a detailed model is not available at present, nor probably, because of the large effort that will be involved, for some time in the future.

We thank Hendrik Schatz for providing details of his calculations of adiabatic expansions and x-ray bursters, and acknowledge communications and conversations with James deBoer, Christian Diget, Hans Fynbo, Alex Heger, Luke Roberts, Hendrik Schatz, Henry Weller, and Michael Wiescher. Research support from: US NSF; grants PHY08-22648 (JINA), PHY-1430152 (JINA-CEE), PHY11-02511.

\begin{thebibliography}{24}
\bibitem[{aus()}]{aus12}
\bibinfo{note}{Sam M. Austin, unpublished compilation; values differ slightly
  from \cite{fre14} because Seeger and Kavanagh, their reference 76, is
  excluded from this calculation and because the input values are not rounded.
  The difference is much smaller than the uncertainties.}
\end{thebibliography}


\end{document}